\documentclass[sigplan,10pt]{acmart}
\pdfoutput=1

\AtBeginDocument{%
  }

\copyrightyear{2024}
\acmYear{2024}
\setcopyright{cc}
\acmConference[PaPoC '24]{The 11th Workshop on Principles and Practice of Consistency for Distributed Data}{April 22, 2024}{Athens, Greece}
\acmBooktitle{The 11th Workshop on Principles and Practice of Consistency for Distributed Data (PaPoC '24), April 22, 2024, Athens, Greece}
\acmDOI{10.1145/3642976.3653029}
\acmISBN{979-8-4007-0544-1/24/04}

\usepackage{pgfplots} 
\usepackage{pgfplotstable}
\usepackage{filecontents}
\usepgfplotslibrary{colorbrewer}
\pgfplotsset{compat=1.16, cycle list/Set1-8} 
\usepackage{tikz}
\usetikzlibrary{
  automata,
  positioning,
  arrows, arrows.meta,
  calc, backgrounds, quotes,
  pgfplots.statistics, pgfplots.colorbrewer,
  patterns
}
\tikzset{
  op/.style={
    font=\footnotesize,
    state,
    minimum size=42.5pt,
    fill=white,
  },
  head/.style={
    op,
    accepting,
  },
  edge/.style={
    ->,
    >={Stealth[round]},
  },
  pred/.style={
    edge,
  },
  anchorref/.style={
    edge,
    blue,
    densely dashed,
  },
  stepmarker/.style={
    font=\footnotesize,
    fill=black!10,
  },
  stepline/.style={
    edge,
    black!70,
    dotted,
    semithick,
    -,
  },
}

\usepackage{amsmath}
\usepackage{algorithm}
\usepackage{algpseudocodex}

\algrenewcommand\algorithmicforall{\textbf{for each}}
\usepackage[noabbrev,nameinlink,capitalize]{cleveref}
\newcounter{figcase}
\renewcommand{\thefigcase}{(\alph{figcase})}
\crefname{figcase}{case}{cases}
\Crefname{figcase}{Case}{Cases}
\creflabelformat{figcase}{#2#1#3}

\usepackage{booktabs}

\usepackage[acronym]{glossaries}
\makeglossaries{}
\newacronym{crdt}{CRDT}{Conflict-Free Replicated Data Type}
\newacronym{mvr}{MVR}{Multi-Valued Replicated Register}
\newacronym{lwwr}{LWWR}{Last-Write-Wins Register}
\newacronym{opid}{OpId}{Operation Id}
\newacronym{lifo}{LIFO}{last-in-first-out}
\newacronym{ot}{OT}{Operational Transformation}

\newcommand*{\fullref}[1]{\hyperref[{#1}]{\autoref*{#1}}}
\newcommand{\op}[3][op]{$\mathit{#1}_{#2}^{#3}$} 
\newcommand{\opid}[2]{$\mathit{_{#1}^{#2}}$}
\newcommand{\setop}[4][set]{$\mathit{#1_{#2}^{#3}(#4)}$}

\newcommand{\undop}[5][undo]{$\mathit{#1_{#2}^{#3}(_{#4}^{#5})}$}
\newcommand{\redop}[5][redo]{$\mathit{#1_{#2}^{#3}(_{#4}^{#5})}$}
\newcommand{\restop}[5][rest]{$\mathit{#1_{#2}^{#3}(_{#4}^{#5})}$}
\newcommand{\stack}[1]{$[$#1$]$}
\def\setabbr{s}
\def\restabbr{rs}
\newcommand{\setopkind}{\textit{SetOp}}
\newcommand{\restopkind}{\textit{RestoreOp}}
\newcommand{\opidtrace}{\textit{OpIdTrace}}

\begin{document}

\title{Undo and Redo Support for Replicated Registers}

\author{Leo Stewen}
\email{lstwn@mailbox.org}
\affiliation{%
  \institution{Technical University of Munich}
  \country{Germany}
}

\author{Martin Kleppmann}
\email{martin@kleppmann.com}
\affiliation{%
 \institution{University of Cambridge}
 \country{United Kingdom}
}


\begin{abstract}
Undo and redo functionality is ubiquitous in collaboration software.
In single user settings, undo and redo are well understood.
However, when multiple users edit a document,
concurrency may arise, leading to a non-linear operation history.
This renders undo and redo more complex
both in terms of their semantics and implementation.

We survey the undo and redo semantics of current mainstream collaboration software
and derive principles for undo and redo behavior in a collaborative setting.
We then apply these principles to a simple \acrshort{crdt}, the \acrlong{mvr},
and present a novel undo and redo algorithm that implements the undo and redo
semantics that we believe are most consistent with users' expectations.
\end{abstract}




\maketitle

\section{Introduction}\label{sec:introduction}

Collaborative editing is an essential feature in modern software.
\glspl*{crdt}~\cite{preguicca2018conflict} are gaining interest because they
enable collaboration in local-first software~\cite{kleppmann2019local},
that works with, but does not require the cloud to function.
They allow for concurrent editing of a shared document without requiring
central coordination while still guaranteeing strong eventual
consistency~\cite{shapiro2011comprehensive}.

As much as collaboration is a common feature today, so is undo and redo support.
Integrating it with \glspl*{crdt} is not trivial.
Although there are some published algorithms for undo and redo in \glspl*{crdt}
(see \cref{sec:related-work}),
their semantics do not match the behavior of current mainstream applications
as detailed in \cref{sec:semantics}.
We believe the semantics of mainstream applications are more in line with
user expectations for an undo feature, and therefore we propose an undo/redo
algorithm for \glspl{crdt} that follows this mainstream behavior.
It works on top of a \gls*{mvr}~\cite{shapiro2011comprehensive},
a core \gls*{crdt} that is often used as a basic
building block to build more complex \glspl*{crdt} via composition.
The intended environment for this algorithm is the Automerge~\cite{automerge} library,
which is a JSON \gls*{crdt} that is based on \glspl*{mvr}.

\section{Semantics of Multi-User Undo/Redo}\label{sec:semantics}

\begin{figure}
\centering
\begin{tikzpicture}[node distance=0.23cm]
  \coordinate (c1) at (0,0);

  \tikzset{
    canvas/.style={
      rectangle,
      draw,
      anchor = west,
      inner sep = 4pt,
    },
    rect/.style={
      rectangle,
      minimum width = 1.2cm, 
      minimum height = 0.5cm,
      inner sep = 0.1cm,
      text = white,
      fill = #1,
    },
    optrans/.style={
      <-,bend right=90,>={Stealth[round]},
      swap,
      black!70,
    },
    next/.style={
      <-,
      >={Stealth[round]},
    },
    caselabel/.style={
      rectangle,
      fill=white,
      minimum width = 238.5pt,
      text width=220pt,
    },
  }

  \def\black{black}
  \def\green{green!60!black}
  \def\red{red}
  \def\upperoffset{16pt}
  \def\casemargin{22pt}
  \def\caselabeldist{1pt}


  \node[
    canvas,
  ] (a1) at (c1) {
    \begin{tikzpicture}[node distance=3pt]
      \node[
          rect=\black,
        ] (r1) at (0,0) {\vphantom{bg}black};
      \node[
          rect=\black,
          below = of r1,
        ] (r2) {\vphantom{bg}black};
    \end{tikzpicture}
  };

  \node[
    canvas,
    right = of a1,
  ] (a2) {
    \begin{tikzpicture}[node distance=3pt]
      \node[
          rect=\red,
        ] (r1) at (0,0) {\vphantom{bg}red};
      \node[
          rect=\black,
          below = of r1,
        ] (r2) {\vphantom{bg}black};
    \end{tikzpicture}
  } edge [next] (a1);

  \node[
    canvas,
    right = of a2,
  ] (a3) {
    \begin{tikzpicture}[node distance=3pt]
      \node[
          rect=\red,
        ] (r1) at (0,0) {\vphantom{bg}red};
      \node[
          rect=\green,
          below = of r1,
        ] (r2) {\vphantom{bg}green};
    \end{tikzpicture}
  } edge [next] (a2);

  \node[
    canvas,
    right = of a3,
  ] (a4) {
    \begin{tikzpicture}[node distance=3pt]
      \node[
          rect=\red,
        ] (r1) at (0,0) {\vphantom{bg}red};
      \node[
          rect=\black,
          below = of r1,
        ] (r2) {\vphantom{bg}black};
    \end{tikzpicture}
  } edge [next] (a3);

  \node[
    canvas,
    right = of a4,
  ] (a5) {
    \begin{tikzpicture}[node distance=3pt]
      \node[
          rect=\red,
        ] (r1) at (0,0) {\vphantom{bg}red};
      \node[
          rect=\green,
          below = of r1,
        ] (r2) {\vphantom{bg}green};
    \end{tikzpicture}
  } edge [next] (a4);


  \node[
    canvas,
    below=\casemargin of a1,
  ] (b1) {
    \begin{tikzpicture}[node distance=3pt]
      \node[
          rect=\black,
        ] (r1) at (0,0) {\vphantom{bg}black};
      \node[
          rect=\black,
          below = of r1,
        ] (r2) {\vphantom{bg}black};
    \end{tikzpicture}
  };

  \node[
    canvas,
    right = of b1,
  ] (b2) {
    \begin{tikzpicture}[node distance=3pt]
      \node[
          rect=\red,
        ] (r1) at (0,0) {\vphantom{bg}red};
      \node[
          rect=\black,
          below = of r1,
        ] (r2) {\vphantom{bg}black};
    \end{tikzpicture}
  } edge [next] (b1);

  \node[
    canvas,
    right = of b2,
  ] (b3) {
    \begin{tikzpicture}[node distance=3pt]
      \node[
          rect=\red,
        ] (r1) at (0,0) {\vphantom{bg}red};
      \node[
          rect=\green,
          below = of r1,
        ] (r2) {\vphantom{bg}green};
    \end{tikzpicture}
  } edge [next] (b2);

  \node[
    canvas,
    right = of b3,
  ] (b4) {
    \begin{tikzpicture}[node distance=3pt]
      \node[
          rect=\black,
        ] (r1) at (0,0) {\vphantom{bg}black};
      \node[
          rect=\green,
          below = of r1,
        ] (r2) {\vphantom{bg}green};
    \end{tikzpicture}
  } edge [next] (b3);

  \node[
    canvas,
    right = of b4,
  ] (b5) {
    \begin{tikzpicture}[node distance=3pt]
      \node[
          rect=\red,
        ] (r1) at (0,0) {\vphantom{bg}red};
      \node[
          rect=\green,
          below = of r1,
        ] (r2) {\vphantom{bg}green};
    \end{tikzpicture}
  } edge [next] (b4);


  \node[
    canvas,
    below=\casemargin of b1,
  ] (3a) {
    \begin{tikzpicture}[node distance=3pt]
      \node[
          rect=\black,
        ] (r1) at (0,0) {\vphantom{bg}black};
    \end{tikzpicture}
  };

  \node[
    canvas,
    right = of 3a,
  ] (c2) {
    \begin{tikzpicture}[node distance=3pt]
      \node[
          rect=\red,
        ] (r1) at (0,0) {\vphantom{bg}red};
    \end{tikzpicture}
  } edge [next] (3a);

  \node[
    canvas,
    right = of c2,
  ] (c3) {
    \begin{tikzpicture}[node distance=3pt]
      \node[
          rect=\green,
        ] (r1) at (0,0) {\vphantom{bg}green};
    \end{tikzpicture}
  } edge [next] (c2);

  \node[
    canvas,
    right = of c3,
  ] (c4) {
    \begin{tikzpicture}[node distance=3pt]
      \node[
          rect=\red,
        ] (r1) at (0,0) {\vphantom{bg}red};
    \end{tikzpicture}
  } edge [next] (c3);

  \node[
    canvas,
    right = of c4,
  ] (c5) {
    \begin{tikzpicture}[node distance=3pt]
      \node[
          rect=\green,
        ] (r1) at (0,0) {\vphantom{bg}green};
    \end{tikzpicture}
  } edge [next] (c4);


  \node[
    canvas,
    below=\casemargin of 3a,
  ] (d1) {
    \begin{tikzpicture}[node distance=3pt]
      \node[
          rect=\black,
        ] (r1) at (0,0) {\vphantom{bg}black};
    \end{tikzpicture}
  };

  \node[
    canvas,
    right = of d1,
  ] (d2) {
    \begin{tikzpicture}[node distance=3pt]
      \node[
          rect=\red,
        ] (r1) at (0,0) {\vphantom{bg}red};
    \end{tikzpicture}
  } edge [next] (d1);

  \node[
    canvas,
    right = of d2,
  ] (d3) {
    \begin{tikzpicture}[node distance=3pt]
      \node[
          rect=\green,
        ] (r1) at (0,0) {\vphantom{bg}green};
    \end{tikzpicture}
  } edge [next] (d2);

  \node[
    canvas,
    right = of d3,
  ] (4d) {
    \begin{tikzpicture}[node distance=3pt]
      \node[
          rect=\black,
        ] (r1) at (0,0) {\vphantom{bg}black};
    \end{tikzpicture}
  } edge [next] (d3);

  \node[
    canvas,
    right = of 4d,
  ] (d4) {
    \begin{tikzpicture}[node distance=3pt]
      \node[
          rect=\green,
        ] (r1) at (0,0) {\vphantom{bg}green};
    \end{tikzpicture}
  } edge [next] (4d);


  \node[
    canvas,
    below=\casemargin of d1,
  ] (e1) {
    \begin{tikzpicture}[node distance=3pt]
      \node[
          rect=\black,
        ] (r1) at (0,0) {\vphantom{bg}black};
    \end{tikzpicture}
  };

  \node[
    canvas,
    right = of e1,
  ] (e2) {
    \begin{tikzpicture}[node distance=3pt]
      \node[
          rect=\red,
        ] (r1) at (0,0) {\vphantom{bg}red};
    \end{tikzpicture}
  } edge [next] (e1);

  \node[
    canvas,
    right = of e2,
  ] (e3) {
    \begin{tikzpicture}[node distance=3pt]
      \node[
          rect=\green,
        ] (r1) at (0,0) {\vphantom{bg}green};
    \end{tikzpicture}
  } edge [next] (e2);

  \node[
    canvas,
    right = of e3,
  ] (e4) {
    \begin{tikzpicture}[node distance=3pt]
      \node[
          rect=\green,
        ] (r1) at (0,0) {\vphantom{bg}green};
    \end{tikzpicture}
  } edge [next] (e3);

  \node[
    canvas,
    right = of e4,
  ] (e5) {
    \begin{tikzpicture}[node distance=3pt]
      \node[
          rect=\green,
        ] (r1) at (0,0) {\vphantom{bg}green};
    \end{tikzpicture}
  } edge [next] (e4);


  \coordinate (d1s) at ($(a1.north)!0.5!(a2.north)+(0,+\upperoffset)$);
  \coordinate (da5) at ($(e1.south)!0.5!(e2.south)+(0,0)$);

  \coordinate (d2s) at ($(a2.north)!0.5!(a3.north)+(0,+\upperoffset)$);
  \coordinate (db5) at ($(e2.south)!0.5!(e3.south)+(0,0)$);

  \coordinate (d3s) at ($(a3.north)!0.5!(a4.north)+(0,+\upperoffset)$);
  \coordinate (dc5) at ($(e3.south)!0.5!(e4.south)+(0,0)$);

  \coordinate (d4s) at ($(a4.north)!0.5!(a5.north)+(0,+\upperoffset)$);
  \coordinate (dd4) at ($(e4.south)!0.5!(e5.south)+(0,0)$);

  \draw (a2.north)+(-0.2cm,+\upperoffset) edge ["A set",optrans] ($(a1.north)+(0,+\upperoffset)$);
  \draw (a3.north)+(-0.2cm,+\upperoffset) edge ["B set",optrans] ($(a2.north)+(0,+\upperoffset)$);
  \draw (a4.north)+(-0.2cm,+\upperoffset) edge ["A undo",optrans] ($(a3.north)+(0,+\upperoffset)$);
  \draw (a5.north)+(-0.2cm,+\upperoffset) edge ["A redo",optrans] ($(a4.north)+(0,+\upperoffset)$);


  \node[caselabel,above=\caselabeldist of a3] {
    \refstepcounter{figcase}\label{fig:two-reg-global}\thefigcase{} Two registers, global undo
  };

  \node[caselabel,above=\caselabeldist of b3] {
    \refstepcounter{figcase}\label{fig:two-reg-local}\thefigcase{} Two registers, local undo
  };

  \node[caselabel,above=\caselabeldist of c3] {
    \refstepcounter{figcase}\label{fig:one-reg-global}\thefigcase{} One register, global undo
  };

  \node[caselabel,above=\caselabeldist of d3] {
    \refstepcounter{figcase}\label{fig:one-reg-local}\thefigcase{} One register, local undo
  };

  \node[caselabel,above=\caselabeldist of e3] {
    \refstepcounter{figcase}\label{fig:one-reg-block}\thefigcase{} One register, remote operation blocks undo
  };
  
\end{tikzpicture}
\caption{
  Different semantics of undo and redo with two users $A$ and $B$ collaboratively
  editing one (two) register(s).
}\label{fig:intro-example}
\end{figure}
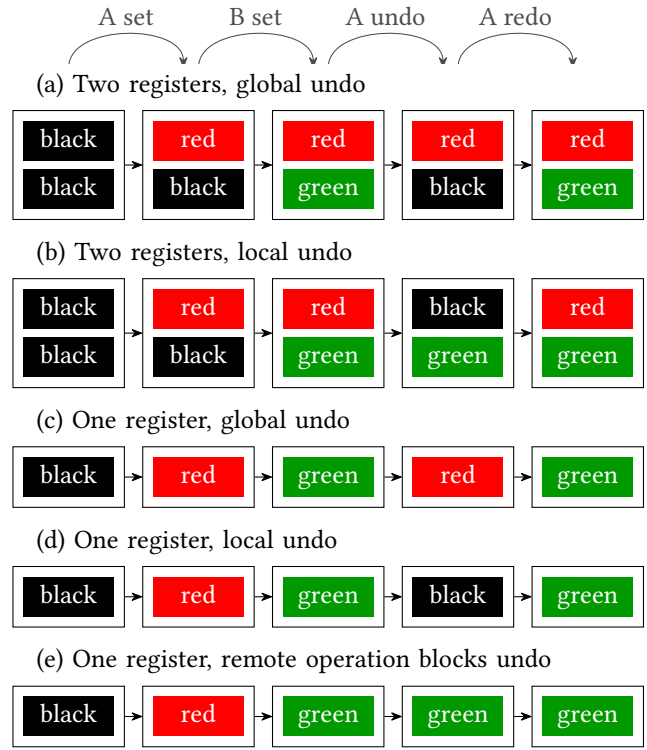

To illustrate the possible semantics of undo and redo with multiple users,
we consider replicated registers that are used to store the fill color
of a rectangle, for instance, on a slide of a presentation deck.
\cref{fig:intro-example} illustrates possible behaviors of undo and redo.
\Cref{fig:two-reg-global,fig:two-reg-local} deal with an application 
with two registers (rectangles) and
\Cref{fig:one-reg-global,fig:one-reg-local,fig:one-reg-block} show an
application with one register. 
We assume that all operations are instantly synced to the other peer and the
operations run sequentially, disregarding concurrency for now.
We distinguish between the following undo behaviors:

\textbf{Global undo.}
In \cref{fig:intro-example}~\ref{fig:two-reg-global},
user $A$ changes the upper rectangle's color
to red, then user $B$ sets the lower rectangle's color to green.
When $A$ subsequently performs an undo, it undoes the most recent operation
\emph{by any user}, i.e., it undoes $B$'s operation by setting the lower
rectangle back to black.
$A$'s redo operation restores the lower rectangle to green.

\Cref{fig:one-reg-global} shows the analogous scenario in which user $A$
and $B$ update the same rectangle.
When $A$ performs an undo, it undoes the most recent operation
again \emph{by any user}, changing the rectangle from green back to red.

\textbf{Local undo.}
In \cref{fig:intro-example}~\ref{fig:two-reg-global},
user $A$ changes the upper rectangle to red,
then user $B$ sets the lower rectangle to green.
Afterwards, user $A$ performs an undo, which undoes the most recent operation
\emph{by $A$ itself}, i.e., changing the upper rectangle back to black.
$A$'s subsequent redo restores the upper rectangle back to red.

In \Cref{fig:one-reg-local}, there is only one rectangle.
First, user $A$ changes it to red, then user $B$ changes it to green.
Subsequently, $A$ performs an undo, and like in the two-register case, this
restores the value of the register to the state before the most recent operation
\emph{by $A$ itself}, i.e., changing it back to black.
When then $A$ performs its redo, the state prior to the undo is restored, i.e.,
green.

\textbf{Remote operation blocks undo.}
In \cref{fig:intro-example}~\ref{fig:one-reg-block}, if user $A$ tries to
perform an undo in a state where the most recent operation is by a different user,
undo is disabled for $A$.

\emph{Local undo} restores to the state prior to the last operation
performed by the same user who issues the undo.
In contrast, \emph{global undo} restores to the state prior the
last operation performed by any user.
However, a redo always restores to the state prior to its corresponding undo.


We tested the behavior of undo and redo in Google Sheets, Google Slides,
Microsoft Excel Online, Microsoft PowerPoint Online, Figma and Miro Boards.
In spreadsheets (Google Sheets and Microsoft Excel Online)
we test a register by setting the value of a single cell instead of recoloring
a rectangle.
All applications we tested exhibit local undo behavior as shown in
\cref{fig:intro-example}~\ref{fig:two-reg-local}~and~\ref{fig:one-reg-local},
except for Miro Boards, which blocks undo every time an operation
by another user on the same register is received, as illustrated
in \cref{fig:intro-example}~\ref{fig:one-reg-block}.
The most common implementations of multi-user undo for a register
can be characterized by the following principles:

\begin{itemize}
  \item \textbf{Local undo}:
    An undo by a user undoes her own last operation on the register,
    thereby possibly also undoing the effects of operations
    by other users,
    if they lie temporally in between her last operation and the time of
    issuing the undo.
  \item \textbf{Undo Redo Neutrality}~\cite{figma2019multiplayer}:
    Assume a register is in state $s$.
    A sequence of $n$ undo operations followed by a sequence of $n$ redo operations
    should restore state $s$.
\end{itemize}

The last principle captures a common usage pattern of undo and redo.
Often users want to look at a past version of a document and then restore
to the most recent version without changing anything.


We think that most applications settled for local undo behavior because
global undo implicitly assumes users being aware of remote changes, too,
while editing a document, and this is a fairly strong assumption to make.
\Cref{fig:intro-example}~\ref{fig:two-reg-global} demonstrates this issue,
assuming that user $A$ has only the upper but not the lower rectangle in her view,
and vice-versa for user $B$.
With global undo, $A$ may be surprised to see no immediate effect of her undo
because it reverted $B$'s green coloring on a different page of the document.
$B$ may be surprised to learn that her last change was undone for no obvious reason.
With more users editing collaboratively, the problem exacerbates, as the
share of remote operations tends to increase.
Hence, we only want to assume that users are aware of their own changes.
Since we think that users favor predictable and consistent undo semantics,
undo behavior with one register should follow local undo behavior as well.

Furthermore, global undo in a \gls{crdt} context suffers from additional problems.
It is possible that multiple users concurrently undo or redo overlapping subsets
of operations, leading to potentially confusing states.
Local undo does not suffer from this problem, since users can only undo 
their own operations.
Finally, global undo presumes that users agree on a total order of operations.
If the total order is determined by (logical) timestamps on operations,
this order cannot be final.
Then, it is possible that a user undoes the operations
it knows in some timestamp interval $[t_1, t_2]$, and subsequently receives
a remote operation with a timestamp $t_{op}$ that falls within the
interval ($t_1 < t_{op} < t_2$).
It is unclear how an algorithm should handle this situation without leading
to further unexpected states.
The fact that most non-\gls{crdt} software chooses local undo over global undo,
despite not being affected by this problem, makes another case for following their lead.

\section{Background on MVRs}\label{sec:background}

A \acrfull{mvr}~\cite{shapiro2011comprehensive} is a data type from the family
of \glspl*{crdt}~\cite{preguicca2018conflict} 
that can be assigned a value, overwriting any previous value.
When multiple values are concurrently assigned, the \gls*{mvr} retains all
values that have not been overwritten.
All values currently held by a register are called its \emph{siblings}
and we require that all replicas return siblings in the same order, e.g.,
sorted by a logical timestamp.
For undo and redo behavior, an undo (or a redo) operation
may reintroduce multiple siblings that have been overwritten in the past.
We assume an operation-based \gls*{crdt} model~\cite{baquero2017pure}.

Replicas store their own copy of the \gls{mvr} and can perform updates on it,
generating an operation which the replica can apply locally,
as well as broadcast to other replicas, which in turn apply it to their own
copy of the \gls{mvr}.
We call an operation \emph{local to a replica} if it originated from that
same replica.
Otherwise, we call it \emph{remote}.

Each operation has a unique identifier (a logical timestamp)
which imposes a total order over all operations.
This order is a linear extension of the happens-before relation~\cite{lamport1978time}.
We call this identifier the \emph{\gls*{opid}}.
An example of such an identifier is pair of a local counter and a unique identifier
for each replica.
We write \op{3}{A} to denote an operation by replica $A$
with a local counter value of $3$.
Whenever a new operation is generated, its counter value is set to one plus
the greatest counter value of any operation known to the generating replica.
This essentially yields a Lamport clock~\cite{lamport1978time}.

When multiple replicas concurrently update the register,
the operation history becomes non-linear.
We model an operation history as a directed acyclic graph
where each node represents an operation and 
each edge from node \op{2}{} to node \op{1}{}
represents a causal dependency of \op{2}{} on \op{1}{}, that is, if
\op{2}{} overwrites the register value previously assigned by \op{1}{}.
We call \op{1}{} a \emph{predecessor} of \op{2}{} and
\op{2}{} a \emph{successor} of \op{1}{}.
Furthermore, we call \op{1}{} an \emph{ancestor} of \op{2}{} and
\op{2}{} a \emph{descendant} of \op{1}{} if there exists a directed
path from \op{2}{} to \op{1}{}.
We call \op{1}{} and \op{2}{} \emph{concurrent} if neither is an ancestor of the
other.
Operations without any successors are called \emph{heads} and
operations without any predecessors are called \emph{roots}.
An operation may have multiple predecessors and multiple successors.
Whenever an operation has multiple predecessors, we call it a \emph{merge} operation.

Operations can be delivered in any order and/or multiple
times because a replica can ensure idempotence by
keeping track of the applied operations through their \glspl*{opid}.
Each replica buffers operations it receives and delivers them
once they are causally ready, i.e., once all ancestors of an operation
have been delivered.

\section{Generating Operations for Undo/Redo}\label{sec:overview}

All operations carry an \gls*{opid} and its set of predecessor \glspl*{opid},
that represent their causal dependencies.
The values produced by these predecessors are the ones that the operation
overwrites.
Next to the \setopkind{}, which sets the \gls{mvr} to the supplied value,
we introduce a second type of operation that we call \restopkind{}.
The payload of this operation (in addition to its \gls*{opid} and predecessors)
is the \gls*{opid} of an ancestor operation which we call its
\emph{anchor} operation.
The effect of this operation is to restore the state of the register to the
state prior to the anchor operation by searching through the operation history.
\restopkind{}s are used to implement both undo and redo.
Moreover, we require additional state besides the operation history.
We introduce two stacks, one for undo and one for redo, with the usual
\acrlong{lifo} semantics.
Both stacks exclusively contain operations generated on the local replica and
ignore remote operations.

When a local \setopkind{} is generated, it is pushed onto the undo stack and
the redo stack is cleared, which causes the loss of the ability to redo after
some sequence of undo operations followed by a \setopkind{}.
This behavior is in line with what most mainstream software does and makes the
undo and redo semantics easier to comprehend.
Clearing the redo stack ensures that redo has a clear next choice of what to redo.
Without clearing the stack, redo (and undo) would become a tree to navigate:
after a sequence of undo operations followed by a \setopkind{}
which is undone again, a subsequent redo would have the choice of redoing
either the previous undo operation or the later \setopkind{}.
This complexity also exists in a single-user setting.
The vim text editor supports undo trees\footnote{
  \url{https://vimhelp.org/undo.txt.html\#undo-branches}
} but it is unusual in this regard.
To avoid the user-facing complexity caused by undo trees,
we follow the mainstream approach of redo stack clearing.

Whenever a user performs an undo, we generate a \restopkind{} that
references the \setopkind{} on top of the undo stack as its anchor
and pop it off the undo stack.
The generated \restopkind{} is pushed onto the redo stack to allow it to be
redone later.
Whenever a user performs a redo, we generate a \restopkind{} that
references the \restopkind{} on top of the redo stack as its anchor and pop
it off the redo stack.
Its anchor is resolved to a \setopkind{} and
pushed onto the undo stack to allow it to be undone later for another time.
For redo operations, its \restopkind{}'s anchor is another \restopkind{},
hence requiring the algorithm to follow one indirection to resolve to a \setopkind{}.

This algorithm ensures that the undo stack contains only \setopkind{}s and
the redo stack contains only \restopkind{}s.
That basically renders a redo as an ``undo of an undo''.

\section{Applying Operations}\label{sec:implementation}

Whenever a replica receives a remote operation,
it buffers it until all its ancestors have been processed,
before adding the new operation to its operation history.
We now introduce the algorithm that determines the current value(s) of the
register given an operation history containing \setopkind{}s and \restopkind{}s.
We split the discussion into three smaller units:
First, \autoref{alg:core-alg} resolves the heads of the operation
history to some \emph{terminal heads} (defined below).
Second, \autoref{alg:gen-values} maps the resulting terminal
heads to the value(s) of the register.
To sort the values in the register appropriately, \autoref{alg:comparison-fn}
defines a comparison function for the terminal heads that \autoref{alg:gen-values}
utilizes.

\newcommand{\var}[1]{\mathit{#1}}

\begin{algorithm}
  \caption{Resolve Heads to Terminal Heads}\label{alg:core-alg}
  \begin{algorithmic}[1]
    \Function{resolveHeads}{$\var{heads}$}
        \State{$\var{todo} \gets ( [\var{head}, ()] \text{ for each } \var{head} \in \var{heads} ) $}
        \State{$\var{termHeads} \gets ()$}
        \While{$\var{todo} \text{ is not empty}$}
          \State{$\var{[nextOp, opIdTrace]} \gets \var{todo.shift()}$}
          \State{$\var{opIdTrace} \gets (\ldots \var{opIdTrace}, \var{nextOp.opId})$}
          \If{isSetOp($\var{nextOp}$)}
            \State{$\var{termHeads.push(\var{[nextOp, opIdTrace]})}$}
          \Else{}
            \Comment{nextOp is a \restopkind{}}
            \State{$\var{anchor} \gets \var{nextOp.anchor}$}
            \ForAll{$\var{pred} \in \var{anchor.predecessors}$}
              \State{$\var{todo.push([pred, opIdTrace])}$}
            \EndFor{}
          \EndIf{}
        \EndWhile{}
        \State{\Return{\Call{getValues}{$\var{termHeads}$}}}
    \EndFunction{}
  \end{algorithmic}
\end{algorithm}

\autoref{alg:core-alg} takes the current set of heads of the operation history
and returns a list of (\setopkind{}, \opidtrace{}) pairs which we
refer to as \emph{terminal heads}.
The \opidtrace{} is a list of \glspl*{opid} from the operations that have been
visited along the path from a head to a \setopkind{}.
We track the \opidtrace{} so that the register's values contributed
by \restopkind{}s can be sorted by the \glspl{opid} of the undo/redo operations
that restore these values,
rather than the \glspl{opid} of the \setopkind{}s that originally assigned
the values to the register.
The \textit{shift()} method on a list pops the first element off the list and
returns it.
The \textit{push()} method on a list appends an element to the list.

The algorithm proceeds as follows:
If a head is a \setopkind{},
it is a terminal head and added to the terminal heads list together with
the \opidtrace{} that contains only the \gls*{opid} of the \setopkind{}.
If a head is a \restopkind{}, the algorithm traverses the operation history
by considering its anchor's predecessors iteratively until \setopkind{}s
are encountered.
While traversing the operation history, every encountered \gls*{opid}
is added to the \opidtrace{} which is passed along to the next iteration.
When the search stops at a \setopkind{}, the \setopkind{} and
the \opidtrace{} are added to the terminal heads list as a pair.
The \opidtrace{}'s last element is always an \gls*{opid} from a \setopkind{}
and any preceding element is an \gls*{opid} from a \restopkind{}.
The processing order of heads and predecessors in the \textit{todo} list
is not relevant for correctness as every terminal head is sorted later in
\cref{alg:gen-values,alg:comparison-fn}.

The reason for handling \setopkind{}s and \restopkind{}s differently 
is that a \restopkind{} may contribute multiple values to
the register due to concurrency in the operation history.
For instance, if at any iteration step a merge \restopkind{} having $k$
predecessors is processed, all $k$ predecessors are considered by the algorithm
and they may produce $k$ or more siblings.

\begin{algorithm}
  \caption{Resolve Terminal Heads to Value(s)}\label{alg:gen-values}
  \begin{algorithmic}[1]
    \Function{getValues}{$\var{termHeads}$}
        \State{$\var{sortedHeads} \gets sort(\var{termHeads}) \text{ desc using Alg.~\ref{alg:comparison-fn}}$}
        \State{$\var{values} \gets ()$}
        \ForAll{$\var{head} \textbf{ in } \var{sortedHeads}$}
          \State{$\var{headOp} \gets \var{head[0]}$}
          \If{$\var{headOp.value} \neq \var{None}$}
            \State{$\var{values.push(headOp.value)}$}
          \EndIf{}
        \EndFor{}
        \State{\Return $\var{values}$}
    \EndFunction{}
  \end{algorithmic}
\end{algorithm}

\autoref{alg:gen-values} determines the value(s) of the register
given a list of terminal heads consisting of pairs of \setopkind{}s and
\opidtrace{}s.
On line 2, it sorts the terminal heads by their \opidtrace{}
in descending order using \cref{alg:comparison-fn}.
Then, it filter maps the sorted terminal heads to their value(s) in lines 3-7.
In case of a deletion (\setopkind{} with no value supplied),
the terminal head is skipped and no value is produced.

\begin{algorithm}
  \caption{Comparison Function}\label{alg:comparison-fn}
  \begin{algorithmic}[1]
    \Function{compare}{$\var{aOpIdTrace}, \var{bOpIdTrace}$}
      \State{$\var{zipped} \gets \mathit{zip}(\var{aOpIdTrace}, \var{bOpIdTrace})$}
      \ForAll{$(aOpId, bOpId) \textbf{ in } \var{zipped}$}
        \If{$aOpId < bOpId$} \Return $\var{Less}$
        \ElsIf{$aOpId > bOpId$} \Return $\var{Greater}$
        \EndIf{}
      \EndFor{}
    \EndFunction{}
  \end{algorithmic}
\end{algorithm}

\Cref{alg:comparison-fn} provides the comparison logic of two \opidtrace{}s
when sorting the list of terminal heads in line 2 of \cref{alg:gen-values}.
The \textit{zip()} function takes two lists and returns a list of pairs
where the first element is from the first list and the second element
is from the second list.
If the lists are of different lengths, the longer list is truncated to the
length of the shorter list.
The for-loop always terminates before finishing with
its last iteration because of the nature of the \opidtrace{}s.
If two terminal heads stem from different heads they already differ in their
first element (the \gls*{opid} of the head).
If two terminal heads stem from the same head (which then must be a \restopkind{}),
they might share a common path but their paths eventually diverge.
The \textit{zip()} function's truncation does not harm this property
as an \opidtrace{}'s last element is always an \gls*{opid} from a \setopkind{}
which cannot occur on the same position of another, \emph{longer} \opidtrace{}.

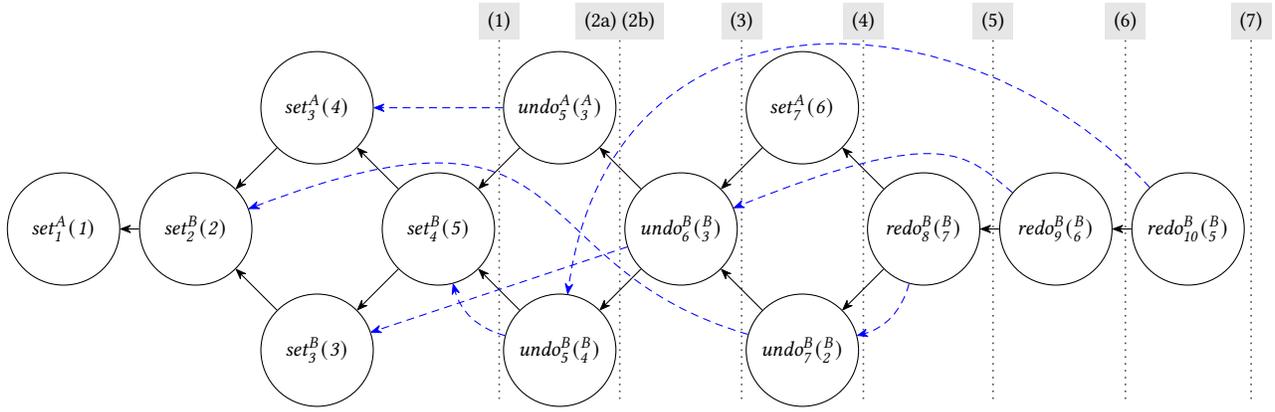
\begin{figure*}
\centering
\begin{tikzpicture}[node distance=50pt]
  \def\dist{22pt}

  \node[op] (a1) at (0, 0) {\setop{1}{A}{1}};
  \node[op,right of=a1] (b2) {\setop{2}{B}{2}} edge [pred] (a1);

  \node[op,above right=\dist of b2] (a3) {\setop{3}{A}{4}} edge [pred] (b2);
  \node[op,below right=\dist of b2] (b3) {\setop{3}{B}{3}} edge [pred] (b2);

  \node[op,below right=\dist of a3] (b4) {\setop{4}{B}{5}} edge [pred] (b3) edge [pred] (a3);

  \node[op,above right=\dist of b4] (a5) {\undop{5}{A}{3}{A}} edge [pred] (b4) edge [anchorref] (a3);
  \node[op,below right=\dist of b4] (b5) {\undop{5}{B}{4}{B}} edge [pred] (b4) edge [anchorref,bend left] (b4);

  \node[op,below right=\dist of a5] (b6) {\undop{6}{B}{3}{B}} edge [pred] (b5) edge [pred] (a5) edge [anchorref] (b3);

  \node[op,above right=\dist of b6] (a7) {\setop{7}{A}{6}} edge [pred] (b6);
  \node[op,below right=\dist of b6] (b7) {\undop{7}{B}{2}{B}} edge [pred] (b6);
  \draw[anchorref] (b7) .. controls (6.0,-0.5) and (7,2.0) .. (b2);

  \node[op,below right=\dist of a7] (b8) {\redop{8}{B}{7}{B}} edge [pred] (b7) edge [pred] (a7) edge [anchorref,bend left] (b7);

  \node[op,right of=b8] (b9) {\redop{9}{B}{6}{B}} edge [pred] (b8);
  \draw[anchorref] (b9) .. controls (12.0,1.0) and (11.7,1.4) .. (b6);

  \node[op,right of=b9] (b10) {\redop{10}{B}{5}{B}} edge [pred] (b9);
  \draw[anchorref] (b10) .. controls (12.0,3.2) and (7.3,3.4) .. (b5);

  \begin{scope}[on background layer]
  \def\divlen{144pt}

  \node[stepmarker,above=50pt of b4,xshift=+23.0pt] (1) {(1)} edge [stepline] ($ (1.center)-(0,\divlen) $);

  \node[stepmarker,right=21.0pt of 1] (2a2b) {(2a) (2b)} edge [stepline] ($ (2a2b.center)-(0,\divlen) $);

  \node[stepmarker,right=21.5pt of 2a2b] (3) {(3)} edge [stepline] ($ (3.center)-(0,\divlen) $);

  \node[stepmarker,right=30.5pt of 3] (4) {(4)} edge [stepline] ($ (4.center)-(0,\divlen) $);

  \node[stepmarker,right=33.5pt of 4] (5) {(5)} edge [stepline] ($ (5.center)-(0,\divlen) $);

  \node[stepmarker,right=34.5pt of 5] (6) {(6)} edge [stepline] ($ (6.center)-(0,\divlen) $);

  \node[stepmarker,right=32.0pt of 6] (7) {(7)} edge [stepline] ($ (7.center)-(0,\divlen) $);

  \end{scope}
\end{tikzpicture}
\caption{
An operation history which is replicated between two replicas $A$ and $B$ with
undo and redo operations. For clarity, undo and redo operations are not labeled
as \restopkind{}s. Their respective anchor operations are indicated by the
dashed blue arrows and denoted in parentheses.
The dotted lines indicate time steps during which we show the nodes' states in
\cref{tab:node-states}.
}\label{fig:op-hist-undo-redo}
\end{figure*}

\newcommand{\ra}[1]{\renewcommand{\arraystretch}{#1}}

\begin{table*}
  \centering
  \small
  \ra{1.3}
  \begin{tabular}{@{}llllllllll@{}}
    \toprule
      Time Step 
        & (1)
        & (2a)
        & (2b)
        & (3)
        & (4)
        & (5)
        & (6)
        & (7)
        \\
    \midrule
      $\mathit{undo Stack}^A$
        & \stack{\setop[\setabbr]{1}{A}{1},\setop[\setabbr]{3}{A}{4}}
        & \stack{\setop[\setabbr]{1}{A}{1}}
        & (2a)
        & (2a)
        & \stack{\setop[\setabbr]{1}{A}{1},\setop[\setabbr]{7}{A}{6}}
        & (4)
        & (4)
        & (4)
        \\
      $\mathit{redo Stack}^A$
        & \stack{}
        & \stack{\restop[\restabbr]{5}{A}{3}{A}}
        & (2a)
        & (2a)
        & \stack{}
        & (4)
        & (4)
        & (4)
        \\
      $\mathit{register}^A$
        & \stack{5}
        & \stack{2}
        & \stack{3,4,2}
        & \stack{2}
        & \stack{1,6}
        & \stack{2}
        & \stack{3,4,2}
        & \stack{5}
        \\
    \midrule
      $\mathit{undo Stack}^B$
        & \stack{\setop[\setabbr]{2}{B}{2},\setop[\setabbr]{3}{B}{3},\setop[\setabbr]{4}{B}{5}}
        & \stack{\setop[\setabbr]{2}{B}{2},\setop[\setabbr]{3}{B}{3}}
        & (2a)
        & \stack{\setop[\setabbr]{2}{B}{2}}
        & \stack{}
        & (3)
        & (2a)
        & (1)
        \\
      $\mathit{redo Stack}^B$
        & \stack{} 
        & \stack{\restop[\restabbr]{5}{B}{4}{B}}
        & (2a)
        & \stack{\restop[\restabbr]{5}{B}{4}{B},\restop[\restabbr]{6}{B}{3}{B}}
        & \stack{\restop[\restabbr]{5}{B}{4}{B},\restop[\restabbr]{6}{B}{3}{B},\restop[\restabbr]{7}{B}{2}{B}}
        & (3)
        & (2a)
        & (1)
        \\
      $\mathit{register}^B$   
        & \stack{5}
        & \stack{3,4}
        & \stack{3,4,2}
        & \stack{2}
        & \stack{1,6}
        & \stack{2}
        & \stack{3,4,2}
        & \stack{5}
        \\
    \bottomrule
  \end{tabular}
\caption{
The internal states of the nodes' undo and redo stacks and the values of the register
at the time steps from \cref{fig:op-hist-undo-redo}.
A \setopkind{} is abbreviated by $\mathit{\setabbr}$ and a \restopkind{}
by $\mathit{\restabbr}$
with its anchor denoted within parentheses.
If a cell contains a reference to another time step, the value is exactly
the same as in that respective step.
The difference between (2a) and (2b) is that in (2a) the replicas have not yet
synced, whereas in (2b) they have synced operations.
At all other time steps the operations are synced.
}\label{tab:node-states}
\end{table*}

\textbf{Example.}\label{sec:algo-example}
\Cref{fig:op-hist-undo-redo} shows an operation history that is replicated
between two replicas $A$ and $B$ that contains undo and redo operations.
\Cref{tab:node-states} shows the internal node states during the time steps
from \cref{fig:op-hist-undo-redo}.
It demonstrates the stack maintenance introduced in \cref{sec:overview} and
shows the register's values produced by
\cref{alg:core-alg,alg:gen-values,alg:comparison-fn} which are applied on top
of the respective operation histories at the time steps.
Steps (2a) and (2b) point out how two concurrent undo operations are resolved.
\undop{5}{A}{3}{A} produces the value $2$ with an \opidtrace{} of
\stack{\opid{5}{A},\opid{2}{B}}.
\undop{5}{B}{4}{B} produces the values $3$ with an \opidtrace{} of
\stack{\opid{5}{B},\opid{3}{B}}
and $4$ with an \opidtrace{} of
\stack{\opid{5}{B},\opid{3}{A}}.
Sorting by these \opidtrace{}s yields the register's values \stack{3,4,2}
after syncing in (2b).

The concurrent \undop{7}{B}{2}{B} and \setop{7}{A}{6} operations at (4) highlight
the necessity of the \opidtrace{}.
The \undop{7}{B}{2}{B} operation produces the value $1$ with an \opidtrace{} of
\stack{\opid{7}{B},\opid{1}{A}}.
The \setop{7}{A}{6} operation immediately yields the value $6$ with an
\opidtrace{} of \stack{\opid{7}{A}}.
If the values were sorted by taking only their \setopkind{}s into account,
the values from a \restopkind{} would always be sorted after
the values from concurrent \setopkind{}s (yielding $[6, 1]$ in the example).
To fix this unfairness, the \opidtrace{} sorts values produced by
\emph{different heads} by their heads' \glspl*{opid}
and values produced by the \emph{same head}
by their first deviation in their \opidtrace{}s (resulting in $[1, 6]$ here).

Finally, at (4) $A$ is giving up its ability to redo because its
\setop{7}{A}{6} operation clears the redo stack,
rendering $A$ unable to redo its previously undone \setop{3}{A}{4} operation.

\textbf{Optimization.}\label{sec:opt}
Note that the predecessors of an operation and the
anchor of a \restopkind{} do not change as further operations are applied.
Hence, we can cache the result of the traversal from a given head
to improve the runtime at the cost of an increased memory overhead.
For each \restopkind{} we cache the values produced by it in sorted order.
Then, we can additionally drop the values'
\opidtrace{}s and rely on a \emph{stable} sort algorithm to include the sorted
values of a \restopkind{} at the right position among the siblings contributed
by other heads, as the cached values' sorting w.r.t.\ the other siblings
is sufficiently determined by the \opidtrace{} starting from a head until the
cached \restopkind{}.

\section{Evaluation}\label{sec:evaluation}

\textbf{Correctness}.
Two replicas that have received the same set of
operations converge to the same state because our algorithm is deterministic,
traverses a finite and cycle-free predecessor relationship,
and does not depend on the order in which operations were received.

Our implementation of the algorithm is written in Typescript and
consists of around 350 lines of code (excluding comments and blank lines),
implementing a \gls*{mvr} with undo and redo.
We test it against a suite of unit tests that covers important
scenarios, both in single-user and multi-user cases.
There are 30 unit tests, comprising over 600 lines of test code.

\textbf{Performance}.
The runtime cost of our algorithm is determined by the graph traversal that
resolves \restopkind{}s at the heads into the corresponding register values.
In case of a \setopkind{} at the head, the runtime of \cref{alg:core-alg}
is constant as the while loop does not add an element, but only
removes the \setopkind{} from the \textit{todo} list.
With caching, the traversal in case of \restopkind{}s is constant, too,
because whenever an anchor's predecessor is a \restopkind{},
its values are already cached in sorted order,
rendering a deeper traversal redundant.
The cost of sorting in \cref{alg:gen-values} is negligible,
as we expect the number of heads to be small for a single register.

\textbf{Integration with Automerge}.
Our implementation is currently a standalone prototype, but we intend to
integrate our algorithm into the Automerge CRDT library in the future.
Since Automerge stores a monotonically growing set of operations,
efficient compression of the serialized operation set is an important property.
Our approach models both undo and redo with a single
operation type whose extra payload consists of just one \gls*{opid}.
Since every undo and redo creates a new \restopkind{} that is added
to the operation set,
the undo and redo stacks are implicitly stored within it and
can be reconstructed upon loading the document to support undo and redo
across sessions without additional persistency overhead.
If an application developer does not require undo and redo functionality,
the value resolution algorithm will be the same as without it,
making library users only pay for what they use.
In terms of space complexity, the additional overhead are
the undo and redo stacks which can be disabled if undo is not required or
bounded if arbitrary undo depth is not wanted.

Finally, we remark an opportunity to prune the history.
Imagine a user undoing a few times and then redoing just as many times
to see the document in the past and then returning to the original state.
We call such a sequence of undo and redo effect-free.
Provided that all replicas have seen the same operations and made no
changes to the register within the effect-free sequence,
the sequence can be pruned (garbage collected) from the history.
However, before doing so causal stability~\cite{baquero2017pure}
must be reached, that is, all
replicas must have sent an operation whose predecessor set contains the
highest \gls*{opid} of the effect-free sequence.

\section{Related Work}\label{sec:related-work}

The terms local and global undo were introduced by the \acrlong{ot}
literature in the context of collaborative text
editing~\cite{sun2000undo,ressel1999reducing, abowd1992giving},
but this distinction is less established in the context of \glspl{crdt}.

\begin{figure}
\centering
\begin{tikzpicture}[node distance=0.23cm]
  \coordinate (c1) at (0,0);

  \tikzset{
    canvas/.style={
      rectangle,
      draw,
      anchor = west,
      inner sep = 4pt,
    },
    rect/.style={
      rectangle,
      minimum width = 1.2cm, 
      minimum height = 0.5cm,
      inner sep = 0.1cm,
      text = white,
      fill = #1,
    },
    optrans/.style={
      <-,bend right=90,>={Stealth[round]},
      swap,
      black!70,
    },
    next/.style={
      <-,
      >={Stealth[round]},
    },
  }

  \def\black{black}
  \def\green{green!60!black}
  \def\red{red}
  \def\upperoffset{2pt}

  
  \node[
    canvas,
  ] (a1) at (c1) {
    \begin{tikzpicture}[node distance=3pt]
      \node[
          rect=\black,
        ] (r1) at (0,0) {\vphantom{bg}black};
    \end{tikzpicture}
  };

  \node[
    canvas,
    right = of a1,
  ] (a2) {
    \begin{tikzpicture}[node distance=3pt]
      \node[
          rect=\red,
        ] (r1) at (0,0) {\vphantom{bg}red};
    \end{tikzpicture}
  } edge [next] (a1);

  \node[
    canvas,
    right = of a2,
  ] (a3) {
    \begin{tikzpicture}[node distance=3pt]
      \node[
          rect=\green,
        ] (r1) at (0,0) {\vphantom{bg}green};
    \end{tikzpicture}
  } edge [next] (a2);


  \node[
    canvas,
    right = of a3,
  ] (a4) {
    \begin{tikzpicture}[node distance=3pt]
      \node[
          rect=\black,
        ] (r1) at (0,0) {\vphantom{bg}black};
    \end{tikzpicture}
  } edge [next] (a3);

  \node[
    canvas,
    right = of a4,
  ] (a5) {
    \begin{tikzpicture}[node distance=3pt]
      \node[
          rect=\red,
        ] (r1) at (0,0) {\vphantom{bg}red};
    \end{tikzpicture}
  } edge [next] (a4);


  \coordinate (d1s) at ($(a1.north)!0.5!(a2.north)+(0,+\upperoffset)$);
  \coordinate (da5) at ($(a1.south)!0.5!(a2.south)+(0,0)$);

  \coordinate (d2s) at ($(a2.north)!0.5!(a3.north)+(0,+\upperoffset)$);
  \coordinate (db5) at ($(a2.south)!0.5!(a3.south)+(0,0)$);

  \coordinate (d3s) at ($(a3.north)!0.5!(a4.north)+(0,+\upperoffset)$);
  \coordinate (dc5) at ($(a3.south)!0.5!(a4.south)+(0,0)$);

  \coordinate (d4s) at ($(a4.north)!0.5!(a5.north)+(0,+\upperoffset)$);
  \coordinate (dd4) at ($(a4.south)!0.5!(a5.south)+(0,0)$);

  \draw (a2.north)+(-0.2cm,+\upperoffset) edge ["A set",optrans] ($(a1.north)+(0,+\upperoffset)$);
  \draw (a3.north)+(-0.2cm,+\upperoffset) edge ["B set",optrans] ($(a2.north)+(0,+\upperoffset)$);
  \draw (a4.north)+(-0.2cm,+\upperoffset) edge ["A undo",optrans] ($(a3.north)+(0,+\upperoffset)$);
  \draw (a5.north)+(-0.2cm,+\upperoffset) edge ["B undo",optrans] ($(a4.north)+(0,+\upperoffset)$);

\end{tikzpicture}
\caption{
  A difficult scenario for counter-based approaches.
}\label{fig:anti-counter}
\end{figure}
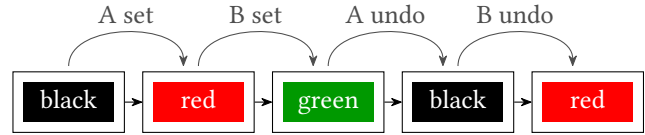

A common approach to undo and redo in the \gls{crdt} literature is
to attach a counter
(sometimes called \emph{degree}~\cite{Weiss2010LogootUndo} or
\emph{undo length}~\cite{Brattli2021undo,Yu2019undo}) to each operation.
Some approaches~\cite{Weiss2010LogootUndo,Martin2010xml} initialize the
counter to $1$ and decrement it on each undo and increment it on each redo.
If the counter is positive, the operation is visible,
otherwise it is invisible.
Other approaches~\cite{Brattli2021undo,Yu2019undo} initialize the counter
to $0$ and increment it on each undo and redo.
If it is even, the operation is visible, otherwise it is not.
When merging conflicting counters of an operation, the maximum is used.
Yet, counter-based approaches handle undo only for a single operation
and do not allow skipping over remote operations, which is required
for local undo as outlined in \cref{sec:semantics}.
\Cref{fig:anti-counter} resembles the single-register scenario from
\cref{fig:intro-example} but instead of $A$ redoing in the last step,
$B$ issues an undo. It shows the correct outcomes with local undo.
Applying counter-based approaches here poses some challenges.
$A$'s undo would render both $A$'s and $B$'s \setopkind{} invisible to
achieve the black coloring.
$B$'s subsequent undo requires to turn $A$'s \setopkind{} visible again to
obtain the red coloring.
To achieve this, $B$'s \emph{undo} must cause the effect of a \emph{redo}
of $A$'s \setopkind{}.
Due to these counterintuitive semantics,
we believe that counter-based approaches are not suitable
for implementing local undo.

Yu et al.~\cite{Yu2015undo} deal with selective undo in the context of strings.
Selective undo is a form of undo which allows a user to undo any operation,
regardless of \emph{where and when} it was generated.
In contrast, global undo allows undoing a remote operation as well
but only in reverse chronological order.
A later work of Yu et al.~\cite{Yu2019undo} describes a generic undo mechanism,
but it only applies to state-based \glspl*{crdt}, assumes global undo
behavior, and focusses on handling concurrent undo and redo of a single operation.
It does not address the issue of (totally) ordering operations on the undo stack
with global undo and a \gls{crdt} context.

Theoretical work by Dolan~\cite{Dolan2020undoable} demonstrates
that any undoable \gls{crdt} is composed of just counters, contradicting our
results at first glance.
Their definition of undoable demands that an undo restores the \gls{crdt}
to a previous state.
We circumvent this by always advancing our internal state and only restoring
the external state.
Martin et al.~\cite{Martin2010xml} discuss undo in the context of XML-like documents.
To the best of our knowledge, Brattli et al.~\cite{Brattli2021undo} is the
only work that also discusses undo and redo in the context of replicated registers,
but they assume global undo behavior.

Yjs~\cite{yjspaper,yjsproject} is another popular \gls{crdt} library with
undo support\footnote{\url{https://docs.yjs.dev/api/undo-manager}}.
Since it uses \acrlongpl{lwwr} instead of \glspl{mvr}, concurrent updates are lost.
Consequently, an undo in Yjs does not recover siblings unlike our approach.
Per default, Yjs implements global undo.
While it allows the changes by a selected replica to be undone
by filtering for ``origins'',
it cannot be used for local undo, as the ability to undo is blocked
upon receiving a change from a non-filtered
origin\footnote{See \Cref{sec:repro} for documentation of this result.}.
Yjs's implementation uses inverse operations instead of pointers into the
operation history.

\section{Conclusion}\label{sec:conclusion}

We tested the semantics of undo and redo of existing collaboration software,
and found that almost all use local undo.
We then presented a novel algorithm for local undo on an \gls*{mvr}
by traversing operation histories represented as directed acyclic graphs.
We believe that our algorithm can also be generalized from a single \gls{mvr}
to more complex \glspl{crdt}, e.g., by treating every key in a map and
every element in a list as a \gls{mvr}, and by using shared
undo and redo stacks for all \glspl{mvr} in the entire data structure.

\begin{acks}
This work was done while Martin Kleppmann was at the Technical University of Munich,
funded by the Volkswagen Foundation and crowdfunding supporters including Mintter
and SoftwareMill.
\end{acks}

\bibliographystyle{ACM-Reference-Format}
\bibliography{references}

\appendix

\section{Reproducibility}\label{sec:repro}

We provide all artifacts of this research on GitHub\footnote{
  \url{https://github.com/lstwn/undo-redo-replicated-registers/tree/papoc-camera-ready}
}.
This gives access to our prototype implementation, its test suite,
\LaTeX{} sources, and documents the surveys of other software
we have conducted.
Yjs' undo behavior is recorded with unit tests.

\section{Experimental Evaluation}

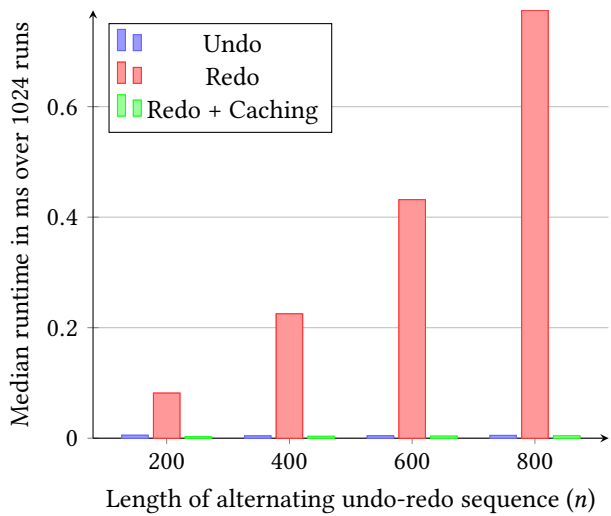
\begin{figure}[b]
\begin{tikzpicture}
\begin{axis}[
  ybar,
  ylabel={Median runtime in ms over 1024 runs},
  axis y line=left,
  ymin = 0,
  xlabel={Length of alternating undo-redo sequence ($n$)},
  axis x line=bottom,
  symbolic x coords={200, 400, 600, 800},
  ymajorgrids=true,
  yminorgrids=true,
  enlarge x limits = 0.2,
  legend pos = north west,
] 

\addplot+[ybar, blue!60, fill=blue!40, postaction={}] coordinates {
(200, 0.005376994609832764)
(400, 0.004238009452819824)
(600, 0.004424989223480225)
(800, 0.005043983459472656)
};

\addplot+[ybar, red!80, fill=red!40, postaction={}] coordinates {
(200, 0.08157795667648315)
(400, 0.22508305311203003)
(600, 0.43157899379730225)
(800, 0.7744320034980774)
}; 

\addplot+[ybar, green!90, fill=green!40, postaction={}] coordinates {
(200, 0.002785980701446533)
(400, 0.00341796875)
(600, 0.003712952136993408)
(800, 0.004239976406097412)
};

\legend{Undo, Redo, Redo + Caching};
\end{axis} 
\end{tikzpicture} 
\caption{
  Runtime of the last undo/redo operation
  in a sequence of alternating undo-redo operations of length $n$
  (counting one undo-redo-pair as one).
}\label{fig:runtime-undo-redo-alt}
\end{figure}

To support the theoretical runtime analysis from \cref{sec:evaluation}
we benchmarked our reference implementation.
The benchmarking was conducted on a 2019 MacBook Pro with a 
2.6 GHz 6-Core Intel Core i7 processor and Node version 18.19.1.

\Cref{fig:runtime-undo-redo-alt} demonstrates that caching preserves
the constant runtime of the algorithm even in the worst case scenario of
resolving a redo of an alternating undo-redo sequence, or more general,
any operation history that causes anchors' predecessors to repeatedly be
\restopkind{}s themselves.
A length $n$ of an alternating undo-redo sequence translates into
a linear operation history of one \setopkind{} (to enable undoing),
followed by $n$ pairs of an undo and a redo.
The figure shows the median runtime to resolve the head to the
register's values over 1024 runs for such sequences of
lengths 200, 400, 600, and 800, if the head is

\begin{itemize}
  \item a \restopkind{} from an undo (blue, first bar),
  \item a \restopkind{} from a redo (red, middle bar),
  \item and a \restopkind{} from a redo but with caching (green, last bar).
\end{itemize}

The findings show that without caching the algorithm's runtime is linear
w.r.t. $n$.
Enabling caching turns the algorithm into a constant time algorithm,
irrespective of the length of the operation history.
Furthermore, it indicates that if space is more precious than runtime,
the price to pay for the traversal without caching may be acceptable,
as even for alternating undo-redo sequences of length 800,
which we believe to be rare in practice, the runtime is below a millisecond.

On the other hand, resolving a non-linear operation history (one which had
concurrent edits) can translate into resolving multiple, albeit shorter,
linear operation histories, in case some encountered \restopkind{}'s anchor
is a merge operation.
If, however, the number of predecessors of an anchor is assumed to be bounded,
e.g., due to a limited number of collaborators on the \gls{mvr},
the constant runtime property remains, as long as caching is used.

\end{document}